\documentclass[aps,prd,twocolumn,nofootinbib,amsmath,amssymb]{revtex4}
\usepackage[dvips]{graphicx}

\newcommand{\tr}{\mathop{\rm Tr}\nolimits}
\def\et{\tilde E}

\begin{document}

\title{Noncommutative quantum mechanics in the presence of delta-function potentials}

\date{\today}

\author{Alexandr Yelnikov}
\affiliation{Department of Physics, Queens College of the CUNY, Flushing, NY 11367 }
\email{yelnikov@qc.edu}

\begin{abstract}
Quantum mechanics in the presence of $\delta$-function potentials is known to be plagued by UV divergencies which result from 
the singular nature of the potentials in question. 
The standard method for dealing with these divergencies is by constructing 
self-adjoint extensions of the corresponding Hamiltonians. Two particularly interesting examples of this kind are nonrelativistic spin zero particles 
in $\delta$-function potential and Dirac particles in Aharonov-Bohm magnetic background. In this paper we show that by extending the corresponding
Schr\"odinger and Dirac equations onto the flat noncommutative space a well-defined quantum theory can be obtained.
Using a star product and Fock space formalisms we construct the complete sets of eigenfunctions and eigenvalues in both cases which turn out to be finite.
\end{abstract}

\maketitle

%%%%%%%%%%%%%%%%%%%%%%%%%%%%%%%%%%%%%%%%%%%%%%%%%%%%%%%%%%%%%%%%%%%%%%%%%%%%%%%%%%%%%%%%%%%%%%%%%%%%%%%%%%%%%%%%%%%%%%
%%%%%%%%%%%%%%%%%%%%%%%%%%%%%%%%%%%%%%%%%%%%%%%%%%%%%%%%%%%%%%%%%%%%%%%%%%%%%%%%%%%%%%%%%%%%%%%%%%%%%%%%%%%%%%%%%%%%%%
\section{Introduction}
Singular interaction potentials were introduced in quantum mechanics more than sixty years ago \cite{old}. Since then they 
found applications in various areas of solid state \cite{condensed}, particle \cite{particle} and nuclear \cite{nuclear} physics.  
It has also been known for a long time that local nature of these potentials at all scales leads to appearance ultraviolet divergencies in quantum mechanics 
similar to those encountered in quantum field theory. However, unlike quantum field theory in which one meets UV divergensies primarilly
due to the presence of infinite number of degrees of freedom, in quantum mechanics the infinities occur due to the singular nature of the potentials chosen. 
Mathematical theory of how to treat such potentials is well known \cite{Reed}, \cite{Alberverio} and is based on construction of self-adjoint 
extensions of the Hamiltonians in question. 
 
On the other hand there has been a lot of recent interest in noncommutative spaces \cite{Madore}, \cite{Connes}, \cite{Konechny} motivated primarilly by string \cite{string} and field theory \cite{field} applications.
In particular, on the field theory side it is believed that in certain cases noncommutative modification
of the algebraic structure of space-time can provide a natural regularization of UV divergencies. This is certainly the case for compact manifolds~\cite{Grosse1}, \cite{Grosse} since field theories  on the noncommutative generalizations of such manifolds 
posess only finite number of degrees of freedom thus removing the intrinsic reason for appearance of UV divergencies.

The purpose of the present paper is to study the quantum mechanics on the flat noncommutative two dimensional space \cite{Nair} in the presence of singular
potentials and to show that nonlocality of interaction induced by fuzziness of space leads to a well-defined quantum theory. In Section~\ref{sec2}  we consider 
nonrelativistic spin zero particles in a $\delta$-function potential and after a brief discussion of the main results known from the commutative case 
\cite{Jackiw}, \cite{Perry}
we give a complete analytic solution of the corresponding Schr\"odinger equation over the noncommutative plane. Section~\ref{sec3} deals
with spin-$\frac{1}{2}$ relativistic particles in the Aharonov-Bohm background magnetic field and we use Fock space 
formalism to obtain the complete set of eigenfunctions for this problem. Also the relationship between commutative and noncommutative solutions is discussed
and we show that in the limit of vanishing noncommutativity parameter $\theta$ we recover the same solutions as those given by theory of self-adjoint extensions.

Our notations and conventions as well as a review of star-product and Fock space formalisms in noncommutative geometry are given in Appendix.

%%%%%%%%%%%%%%%%%%%%%%%%%%%%%%%%%%%%%%%%%%%%%%%%%%%%%%%%%%%%%%%%%%%%%%%%%%%%%%%%%%%%%%%%%%%%%%%%%%%%%%%%%%%%%%%%%%%%%%%%%%
%%%%%%%%%%%%%%%%%%%%%%%%%%%%%%%%%%%%%%%%%%%%%%%%%%%%%%%%%%%%%%%%%%%%%%%%%%%%%%%%%%%%%%%%%%%%%%%%%%%%%%%%%%%%%%%%%%%%%%%%%%
\section{\label{sec2} Two-Dimensional Delta-function Potential.}

\subsection{Commutative case.}

In the commutative case the Schr\"odinger equation for spin zero particle moving in a two-dimensional $\delta$-function
potential, can be written as ($\hbar = 1$)
\begin{equation}
-\frac{1}{2m} \nabla^2 \Psi({\bf r}) +V_0 \delta ({\bf r}) \Psi({\bf r}) = E \Psi({\bf r})
\end{equation}
or, in terms of new coupling constant $\alpha_0 = 2mV_0$ and energy parameter $\et = 2mE$,
\begin{equation}
-\nabla^2 \Psi({\bf r}) +\alpha_0 \delta ({\bf r}) \Psi({\bf r}) = \et \Psi({\bf r}). \label{commequation}
\end{equation}
Both the $\delta$-function potential in two dimensions and the kinetic energy operator scale in polar coordinates $(\rho, \phi)$
as $1/\rho^2$,
therefore the coupling $\alpha_0$ is dimensionless. As a consequence, the Hamiltonian is scale invariant and 
we can anticipate the presence of logarithmic ultraviolet divergencies, analogous to those appearing in QED and QCD.
The standard way of obtaining an exact solution of (\ref{commequation}) analytically is to use method of self-adjoint extensions.
As explained in \cite{Jackiw}, the two-dimensional Laplace operator $\nabla^2$ is not self-adjoint on a punctured plane and construction
of self-adjoint extension requires relaxing the condition of regularity of wave functions at the origin and allowing $\log \rho$
singularity at $\rho = 0$. However, $\delta(\rho)\log \rho$ is then not well-defined. Thus, we need to define $\delta(\rho)\Psi(\rho, \phi)$ for
wavefunctions behaving near the origin as 
\begin{equation}
\Psi(\rho, \phi)= \psi_0 \log(\mu\rho) + \psi_1 + O(\rho).
\end{equation}
where $\mu$ is an arbitrary dimensional parameter. Formal integration of eq.(\ref{commequation}) over a small disk of radius $\epsilon$
followed by taking the limit $\epsilon \to 0$ gives the correct constraint on coefficients
$\psi_0$ and $\psi_1$
\begin{equation}
2\pi\psi_0 - \alpha_0 \psi_1 = 0 \label{boundarycond}
\end{equation}
which we should take as a boundary condition on a wavefunction. More detailed derivation of (\ref{boundarycond}) can be found in \cite{Reed}.

With these ideas in mind we can rewrite (\ref{commequation}) for an axially symmetric wavefunction as 
\begin{equation}
\Psi^{\prime\prime}(\rho) + \frac{1}{\rho}\Psi^{\prime}(\rho) + \et\Psi(\rho) = 0
\end{equation}
which for $\et < 0$ admits
\begin{equation}
\Psi(\rho) = K_0\left(\sqrt{|\et|}\ \rho\right)
\end{equation}
as a solution. Here $K_{0}(x)$ is the modified Bessel function of the third kind \cite{Bateman}. From the asymptotic behaviour of $\Psi(\rho)$ near the origin, namely,
\begin{equation}
\Psi(\rho) = -\log\left(\frac{\et \rho}{2}\right) - \gamma + O(\rho)
\end{equation} 
and the boundary condition (\ref{boundarycond}) we find an expression for the energy of this bound state
\begin{equation}
\et = -4\mu^2 e^{-2\gamma}e^{\frac{4\pi}{\alpha_0}}. \label{eigenval}
\end{equation}
Therefore this theory provides us with an example of a spontaneous breakedown of scale invariance where bound state energy is
set by an arbitrary dimensionful parameter $\mu$. 

%%%%%%%%%%%%%%%%%%%%%%%%%%%%%%%%%%%%%%%%%%%%%%%%%%%%%%%%%%%%%%%%%%%%%%%%%%%%%%%%%%%%%%%%%%%%%%
%%%%%%%%%%%%%%%%%%%%%%%%%%%%%%%%%%%%%%%%%%%%%%%%%%%%%%%%%%%%%%%%%%%%%%%%%%%%%%%%%%%%%%%%%%%%%%
\subsection{Noncommutative case}

In complete analogy with commutative case we can write the Schr\"odinger equation for a particle moving in a
noncommutative $\delta$-function potential as
\begin{equation}\label{nmain}
-\hat\nabla^2 \hat\Psi({\bf\hat x}) + \alpha_0\hat \delta(\hat x) \hat\Psi({\bf\hat x}) = \et \hat\Psi({\bf\hat x}), 
\end{equation}
but where $\hat\Psi({\bf\hat x})$ now is an operator valued  wavefunction
$$
\hat\Psi({\bf\hat x}) = \int d^2p\  \Phi(p, \bar p) e^{i(pz +\bar p\bar z)}
$$
and $\hat\delta(\hat x)$ is defined in (\ref{Nelta}).

In momentum space equation (\ref{nmain}) becomes
\begin{equation}\label{main}
(p\bar p -\et)\Phi(p, \bar p) = - \frac{\alpha_0}{(2\pi)^2}\int d^2 \eta\  \Phi(\eta, \bar \eta)\  e^{-\frac{\theta}{4}|p-\eta|^2+\frac{\theta}
{4}(\bar p\eta - p\bar \eta)}
\end{equation}
with parameters $\et = 2mE$ and $\alpha_0 = 2mV_0$ defined as in commutative case.

Equation (\ref{main}) is an integral equation for momentum space wave-function which, in general, is difficult to solve. But
the solution of this particular equation is greatly simplified if one observes that the differential operator
\begin{equation}
\frac{\partial}{\partial p} + \frac{\theta}{4}\bar p
\end{equation}
when applied to the r.h.s of (\ref{main}) gives zero, and, therefore, we are left only with
\begin{equation}
\left[\frac{\partial}{\partial p} + \frac{\theta}{4}\bar p \right]  (|p|^2 -\et)\Phi(p, \bar p) = 0 \label{simpleq}
\end{equation}
This last equation is easy to solve and, after we pass to the eigenstates $\Phi_n(p)$ of angular momentum operator 
\begin{equation}
\Phi(p, \phi) = \sum_{n=-\infty}^{\infty}\ \Phi_n(p) e^{in\phi},
\end{equation}
its solutions are given by
\begin{equation}\label{sol}
\Phi_n(p)= \frac{p^n}{p^2-\et}\ e^{-\frac{\theta}{4}p^2}, \ \ \ \ n=0, \pm 1, \pm 2, \ldots\ 
\end{equation}
for $\et < 0$, while for positive energies eq.(\ref{simpleq}) admits an extra $\delta$-function term and therefore we write
\begin{equation}\label{possol}
\Phi_n(p)= \delta(p^2 - \et) + C_n \frac{p^n}{p^2-\et}\ e^{-\frac{\theta}{4}p^2}
\end{equation}
By direct substitution, it can be verified that eq.(\ref{possol}) gives a complete set of scattering states provided that constants $C_n$
are choosen as the solutions of
\begin{equation}
1 = -\frac{\alpha_0}{4 \pi}\left\{C_n I_n\left(\frac{\theta \et_n}{2}\right) +{1\over{n!}}\ e^{-{\theta\over{4}}\et}
\left( {\theta\over{2}}\sqrt{\et}\right)^n \right\}
\end{equation}
for $n \ge 0$, and $C_n = 0$ for $n < 0$, with functions $I_n$ defined by
\begin{equation}\nonumber
I_n(a)={1\over{n!}}\int_0^\infty dt \frac{t^n}{t-a} e^{-t}= (-a)^n \Psi(n+1,n+1, -a)
\end{equation}
and where $\Psi(a, b, c)$ is the confluent hypergeometric function.

Similar analysis shows that Eqs.\ref{sol} also satisfy Schr\"odinger equation for $n \ge 0$ only, thus giving us bound state
solutions the number of which, as well as their energies being given by the following set of eigenvalue equations
\begin{equation}\label{bound}
1 = -\frac{\alpha_0}{4 \pi}\ I_n\left(-\frac{\theta |\et_n|}{2}\right).
\end{equation}
\begin{figure}
\begin{center}
\includegraphics[width=8.3cm]{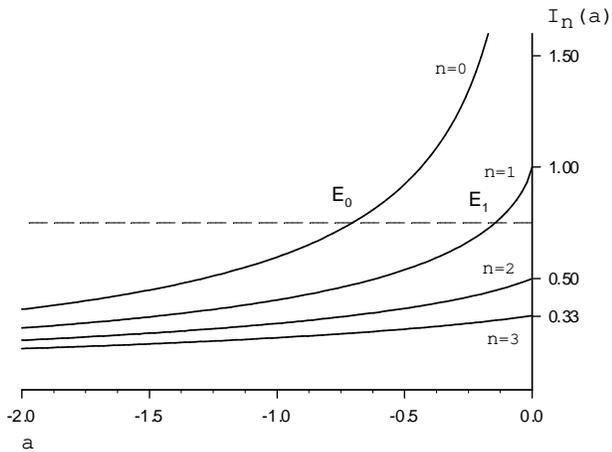}
\caption{Example of graphical solution of eigenvalue equation (\ref{sol}). In this case we have two bound states.}
\end{center}
\end{figure}
Example of a graphical solution of these equations is presented in Fig.1, from which it can also be seen, that like in commutative case, the radially symmetric solution ($n=0$) exists for
arbitrary strength of attractive potential but, unlike the commutative case, for sufficiently strong potentials (when dimensionless 
coupling $|\alpha_0| > 1$)
this problem also admits bound states with nonzero angular momentum ($n>0$), the number of such solutions being a function
of $\alpha_0$ only.  It is also instructive to look at limiting cases of very strong and very weak binding potential more closely.

%%%%%%%%%%%%%%%%%%%%%%%%%%%%%%%%%%%%%%%%%%%%%%%%%%%%%%%%%%%%%%%%%%%%%%%%%%%%%%%%%%%%%%%%%%%%%%%%%%%%%
%%%%%%%%%%%%%%%%%%%%%%%%%%%%%%%%%%%%%%%%%%%%%%%%%%%%%%%%%%%%%%%%%%%%%%%%%%%%%%%%%%%%%%%%%%%%%%%%%%%%%
\subsubsection{Large coupling ($|\alpha|\gg 1$)}

\noindent In this case $a=\frac{\theta |\et|}{2}\gg 1$ and we can use asymptotic expansion
$$
I_n(a) = {1\over a} - {\frac{n+1}{a^2}} + O\left({1\over{a^3}}\right) 
$$ 
and write (\ref{sol}) as
$$
1 = \frac{|\alpha_0|}{4\pi}\left( {1\over {a_n}} - {\frac{n+1}{a_2^2}} \right)
$$
with solutions
$$
a_n = \frac{4\pi}{|\alpha_0|} - (n+1)
$$
or
$$
\et_n = -\frac{2}{\theta}a_n = \frac{8\pi}{\theta\alpha_0} + \frac{2}{\theta}(n+1)
$$
and we see that bound state spectrum in this case coincides with that of a simple harmonic ocsillator of frequency
$\frac{2}{\theta}$ \cite{Gamboa}.

\subsubsection{Small coupling ($|\alpha|\ll 1$)}

In this  case only one solution of (\ref{bound}) exists corresponding to $n=0$ and $a=\frac{\theta |\et|}{2}\ll 1$.
Therefore, we can use the asymptotic form of $I_n(a)$ for small $a$'s which is
\begin{eqnarray}
I_0(a) = -\ln |a| - \gamma + O(|a|\ln|a|)\\
\gamma = 0.5772\ldots\ ,\ \mbox{Euler's constant}
\end{eqnarray}
This gives the following eigenvalue equation
$$
1=\frac{|\alpha_0|}{4\pi}(\ln|a| + \gamma) 
$$
with binding energy
$$
|\et| = \frac{2}{\theta} e^{-\gamma}e^{\frac{4\pi}{\alpha_0}}
$$
Comparing this last expression with (\ref{eigenval}) we see that this limit corresponds to the commutative case with $1\over\theta$
playing the r\^ole of parameter $\mu^2$ in commutative case.

%%%%%%%%%%%%%%%%%%%%%%%%%%%%%%%%%%%%%%%%%%%%%%%%%%%%%%%%%%%%%%%%%%%%%%%%%%%%%%%%%%%%%%%%%%%%%%%%%%%%%%%%%%%%%%%%%%%%%%%%%%
%%%%%%%%%%%%%%%%%%%%%%%%%%%%%%%%%%%%%%%%%%%%%%%%%%%%%%%%%%%%%%%%%%%%%%%%%%%%%%%%%%%%%%%%%%%%%%%%%%%%%%%%%%%%%%%%%%%%%%%%%%
\section{\label{sec3} Fermions in a magnetic vortex background}

\subsection{Commutative case}
In this section we briefly review the solutions of a massive Dirac equation in the Aharonov-Bohm background field of an infinitely thin
magnetic vortex carrying magnetic flux $\Phi$. This presentation closely follows Ref.\cite{Gerbert}. The electromagnetic potential describing such field configuration can be chosen
in polar coordinates $(\rho, \phi)$ as ${\mathbf A} = -\frac{e\Phi}{2\pi}\frac{1}{\rho}{\mathbf{ \hat{\phi}}}$. $\mathbf A$ has a well known property of being locally a pure gauge.

The Dirac equation for this problem is
\begin{equation}\label{Dirac}
(i\not{\!\partial} + \not{\!\!A} -m)\Psi(t, r, \phi) =0.
\end{equation}
and allows passing to the eigenstates of angular momentum $n+\frac{1}{2}$. By defining
\begin{equation}
\Psi_{E,n}(t,\rho,\phi) = 
\left(
\begin{array}{c}
\chi^1(\rho) e^{in\phi} \\
\chi^2(\rho) e^{i(n+1)\phi} 
\end{array} 
\right)
e^{-iEt}
\end{equation}
the radial eigenvalue problem is
\begin{equation}
\left(
\begin{array}{cc}
m - E & -i\left( \partial_{\rho} + \frac{\nu+1}{\rho}\right) \\
-i\left(\partial_{\rho} - \frac{\nu}{\rho}\right) & -m - E
\end{array}
\right) \chi(\rho) = 0
\end{equation}
with $\nu \equiv n+\Phi$. For $E^2 > m^2$ it has the solutions
\begin{equation}
\chi_{\nu}(\rho) =\frac{1}{N}
\left(
\begin{array}{c}
\sqrt{E+m}(\epsilon_n)^n J_{\epsilon_n \nu}(k\rho) \\
i\sqrt{E-m}(\epsilon_n)^{n+1} J_{\epsilon_n (\nu+1)}(k\rho)
\end{array}
\right)
\end{equation}
where $N$ is a normalization factor, $k = \sqrt{E^2-m^2}$, $J_{\lambda}(x)$ denotes the Bessel functions and $\epsilon_n$ is taken to be
either$+ 1$ or $-1$ to assure regularity of spinor components at the origin. This choice of the sign for $\epsilon_n$ can always be done except for the partial wave
with
\begin{equation}
-1 < \nu <0
\end{equation}  
in which case both choices of sign lead to solutions that are square integrable, though singular in one component, at the origin.
To avoid a loss of completeness in angular basis, a family of self-adjoint extensions of Dirac Hamiltonian is required. These extensions are 
parametrized by a single parameter $0 \le \Theta \le 2\pi$ (not to be confused with noncommutativity parameter $\theta$) and restrict the behaviour
of the wavefunction at $\rho \to 0$ to be
\begin{equation}
\lim_{\rho\to 0}\chi(\rho) \sim
\left(
\begin{array}{c}
i \rho^{\nu} \sin\left(\frac{\pi}{4}+\frac{\Theta}{4} \right) \\
\rho^{-\nu-1} \cos\left(\frac{\pi}{4}+\frac{\Theta}{4} \right)
\end{array}
\right)
\end{equation}

With the boundary condition established, the energy eigenstates are
\begin{widetext}
\begin{equation}\label{extensions}
\chi_{\nu}(\rho)\sim 
\left(
\begin{array}{c}
\sqrt{E+m}\ [\sin\mu J_{\nu}(k\rho) +(-1)^n \cos\mu J_{-\nu}(k\rho)] \\
i\sqrt{E-m}\ [\sin\mu J_{\nu+1}(k\rho) +(-1)^{n+1} \cos\mu J_{-(\nu+1)}(k\rho)] \\
\end{array}
\right)
\end{equation}
\end{widetext}
with $\mu$ related to $\Theta$ by the equation
\begin{equation}
\begin{array}{rl}
\tan\left(\frac{\pi}{4}+\frac{\Theta}{2} \right) = & (-1)^n \left(\frac{E+m}{E-m}\right)^{1/2}\left(\frac{k}{2m}\right)^{2\nu+1}\\
& \times\frac{\Gamma(-\nu)}{\Gamma(\nu+1)}\tan\mu .
\end{array}
\end{equation} 
In addition, for $\pi/2<\Theta<3\pi/2$ there is a bound state
\begin{equation}
B_{\nu}(\rho)\sim 
\left(
\begin{array}{c}
\sqrt{m+E} K_{\nu}(\bar k\rho)\\
i\sqrt{m-E} K_{\nu+1}(\bar k\rho)
\end{array}
\right)
\end{equation}
where $\bar k = -ik = \sqrt{m^2-E^2}$ and $K_{\nu}(x)$ are modified Bessel functions. The bound-state energy is implicitly determined
from
\begin{equation}
\frac{(1+E/m)^{\nu+1}}{(1-E/m)^{-\nu}} = -2^{2\nu+1}\frac{\Gamma(\nu+1)}{\Gamma(-\nu)}\tan\left(\frac{\pi}{4}+\frac{\theta}{2}\right).
\end{equation}

%%%%%%%%%%%%%%%%%%%%%%%%%%%%%%%%%%%%%%%%%%%%%%%%%%%%%%%%%%%%%%%%%%%%%%%%%%%%%%%%%%%%%%%%%%%%%%%%%%%%%%%%
%%%%%%%%%%%%%%%%%%%%%%%%%%%%%%%%%%%%%%%%%%%%%%%%%%%%%%%%%%%%%%%%%%%%%%%%%%%%%%%%%%%%%%%%%%%%%%%%%%%%%%%%
\subsection{Noncommutative case}

On the noncommutative plane it is convenient to use complex notation for the vector potential
\begin{equation}
A_z = \frac{1}{2}(A_1 - i A_2), \ \ \ \ \ A_{\bar z} = \frac{1}{2}(A_1 + i A_2), 
\end{equation} 
so that magnetic field strength can be written as
\begin{equation}
F_{z \bar z} = \partial_z A_{\bar z} - \partial_{\bar z} A_z -ie[A_z,A_{\bar z}] .
\end{equation}
For a magnetic vortex field
\begin{equation}
B = 2i F_{z \bar z} = \frac{\Phi}{2\pi\theta}|0\rangle \langle 0|
\end{equation}
an explicit expression for vector potential can be written if we use the following ansatz
\begin{equation}
A_{z} = i\frac{\bar z}{2\theta}\Delta, \ \ \ \ A_{\bar z} = -i \Delta \frac{z}{2\theta}, 
\end{equation}
with a radially symmetric function
\begin{equation}
\Delta = \frac{1}{e}\sum_{n=0}^{\infty}\left(1-\sqrt{1-\frac{e\Phi}{2\pi}\frac{1}{n+1}}\right) |n\rangle \langle n|
\end{equation}
This form is valid for $1 - \frac{e\Phi}{2\pi} > 0$ only.

The Dirac equation for this problem formally coincides with eq.(\ref{Dirac}) in commutative case
\begin{equation}
(i\not{\!\partial} + \not{\!\!A} -m)\Psi =0.
\end{equation}
if we require that gauge connection acts from the left on operator-valued  spinor wave function $\Psi$.
By defining 
\begin{equation}
\Psi_{E} =\left(
\begin{array}{c}
\chi^1 \\ \chi^2
\end{array}
\right)e^{-iEt}
\end{equation}
we get the 
following eigenvalue problem
\begin{equation}\label{firstorder}
\left(
\begin{array}{cc}
E-m & 2i\left(\frac{\partial}{\partial_z} -ieA_z\right)\\
-2i\left(\frac{\partial}{\partial_{\bar z}} - ieA_{\bar z}\right) & -E-m
\end{array}
\right)
\left(
\begin{array}{c}
\chi^1 \\ \chi^2
\end{array}
\right) = 0
\end{equation}
which leads to the second order equations for each of the two components of the Dirac spinor
\begin{eqnarray}
\left(\frac{\partial}{\partial_z} -ieA_z\right)\left(\frac{\partial}{\partial_{\bar z}} - ieA_{\bar z}\right) \chi^1 = 
-\frac{k^2}{4}\chi^1, \label{eq1} \\
\left(\frac{\partial}{\partial_{\bar z}} - ieA_{\bar z}\right)\left(\frac{\partial}{\partial_z} -ieA_z\right)\chi^2 = 
-\frac{k^2}{4}\chi^2 . \label{eq2}
\end{eqnarray}
It should be noted here that due to the noncommutativity of covariant derivatives
\begin{equation}
\left[
\left(\frac{\partial}{\partial_z} -ieA_z\right), \left(\frac{\partial}{\partial_{\bar z}} - ieA_{\bar z}\right)
\right] = ieF_{z\bar z}
\end{equation}
the equation for $\chi^1$ component contains an extra term $ieF_{z\bar z}\chi^1$ as compared to the equation satisfied by $\chi^2$.
In a commutative limit this term is proportional to $\delta(x)\chi^1(x)$ and is equal to zero as long as $\chi^1$ is regular 
at the origin.

To solve (\ref{eq1}, \ref{eq2}) we use the following ansatz of angular momentum $n+{\frac{1}{2}}$
\begin{equation}
\chi =
\left(
\begin{array}{c}
\sum\limits_{l=0}^{\infty} \chi^1_l |l-n\rangle \langle l |\\
\sum\limits_{l=0}^{\infty} \chi^2_l |l-n-1\rangle \langle l |
\end{array}
\right)
\end{equation}
for $n = -1,-2, \ldots$ and
\begin{equation}\label{positive}
\chi =
\left(
\begin{array}{c}
\sum\limits_{l=n}^{\infty} \chi^1_l |l-n\rangle \langle l |\\
\sum\limits_{l=n+1}^{\infty} \chi^2_l |l-n-1\rangle \langle l |
\end{array}
\right)
\end{equation}
for $n=0,1,2, \ldots$.

For negative $n$'s this gives us recursion relations on coefficients $\chi^{1,2}_l$
\begin{eqnarray}
\begin{array}{rl}
\chi^{1,2}_{l+1}\sqrt{l+1}\sqrt{l-\nu +1}& + \ \chi^{1,2}_{l-1}\sqrt{l}\sqrt{l-\nu} \\ 
- \chi^{1,2}_{l}&\!\!(2l -\nu +1 -\frac{\theta}{2}k^2) =0 \label{recursion}
\end{array}\\
\chi^{1,2}_1 \sqrt{-\nu +1} -\chi^{1,2}_0(-\nu +1 -\frac{\theta}{2}k^2) = 0 \label{boundary}
\end{eqnarray}
where again we used $\nu = n+\Phi$. These recursion relations are quite easy to solve with the solutions given
by (up to a normalization factor)
\begin{subequations}\label{solution1}
\begin{eqnarray}
\chi^1_l = \frac{i(E+m)}{\sqrt{-\nu}}\sqrt{\frac{\theta}{2}}\sqrt{\frac{l!}{(-\nu+1)_l}} L^{-\nu}_{l}\left(\frac{k^2\theta}{2}\right)\\
\chi^2_l = \sqrt{\frac{l!}{(-\nu)_l}} L^{-\nu-1}_{l}\left(\frac{k^2\theta}{2}\right) 
\end{eqnarray}
\end{subequations}
with $(\alpha)_l = \alpha(\alpha+1)\ldots (\alpha +l-1)$ the Pochhammer symbol and $L_l^{\alpha}(x)$ the generalized Laguerre polynomial. 

For $n=0,1,2,\ldots$ recursion relations are the same as (\ref{recursion}) but "boundary" conditions (\ref{boundary}) are different
(note that in this case the series expansion in (\ref{positive}) begins with $\chi^1_n$ and $\chi^2_{n+1}$ terms for the
first and second spinor components respectively) 
%\begin{widetext}
\begin{eqnarray}
\begin{array}{lr}
\chi^1_{n+1}\sqrt{n+1}\sqrt{n-\nu+1} - \chi^1_{n}(n+1-\frac{\theta}{2}k^2) = 0 \label{bound1}&
\end{array}\\
\begin{array}{rl}
\chi^2_{n+2}\sqrt{n+2}\sqrt{n-\nu+1}&\\
- \chi^2_{n+1}&\!\!(2(n+1)-\nu-\frac{\theta}{2}k^2) = 0 \label{bound2}
\end{array}
\end{eqnarray}
%\end{widetext}
These equations are also easy to solve
\begin{widetext}
\begin{equation}\label{solution2}
\chi = \frac{1}{N}
\left(
\begin{array}{c}
\frac{i(E+m)}{\sqrt{-\nu}}\sqrt{\frac{\theta}{2}} \sum\limits_{l=n}^{\infty} \sqrt{\frac{l!}{(-\nu+1)_l}}\left\{
L^{-\nu-1}_l(\frac{k^2\theta}{2}) + \frac{-2\nu A^{\nu}}{k^2\theta}(-\nu+1)_l \Psi(l+1, 1+\nu, -\frac{k^2\theta}{2})\right\}|l-n\rangle\langle l|\\
\sum\limits_{l=n+1}^{\infty} \sqrt{\frac{l!}{(-\nu)_l}}\left\{
L^{-\nu-1}_l(\frac{k^2\theta}{2}) + A^{\nu}(-\nu)_l \Psi(l+1, 2+\nu, -\frac{k^2\theta}{2})\right\}|l-n-1\rangle\langle l|\\
\end{array}
\right)
\end{equation}
\end{widetext}
where coefficient $A^{\nu}$ is a solution of the linear equation
\begin{widetext}
\begin{equation}
(n-\nu)L^{-\nu-1}_n (\frac{k^2\theta}{2}) + A^{\nu}(-\nu)_{n+1} \Psi(n+1, 2+\nu, -\frac{k^2\theta}{2}) = 0
\end{equation}
\end{widetext}
and ensures that conditions (\ref{bound1}, \ref{bound2}) are obeyed. It is an easy  task now to check that eqs.(\ref{solution1}), (\ref{solution2})
do also satisfy the first order Dirac equations (\ref{firstorder}) and, therefore, give a complete set of angular momentum eigenstates for our problem.

%%%%%%%%%%%%%%%%%%%%%%%%%%%%%%%%%%%%%%%%%%%%%%%%%%%%%%%%%%%%%%%%%%%%%%%%%%%%%%%%%%%%%%%%%%%%%%%%%%%%%%%%%%%%%%%%%%%%
%%%%%%%%%%%%%%%%%%%%%%%%%%%%%%%%%%%%%%%%%%%%%%%%%%%%%%%%%%%%%%%%%%%%%%%%%%%%%%%%%%%%%%%%%%%%%%%%%%%%%%%%%%%%%%%%%%%%
\section{Conclusions}

In this paper we have studied noncommutative generalizations of quantum mechanics in the presence of $\delta$~-~function potentials.
It was found that noncommutativity of space-time can be used to provide an intrinsic regularization of the theories in question. 
Using the star product formalism we found analytically all the solutions of that problem .
The folowing remarks, however, are in order:
\begin{enumerate}
\item
The apparent asymmetry between holomorphic and antiholomorphic solutions in, for example (\ref{possol}), can be understood if one notes that action of noncommutative 
$\delta$-function operator on antiholomorphic wavefunctions is trivial
\begin{equation}
\hat \delta(\hat x)\hat\Psi = 0
\end{equation}
and, therefore, in our model these modes are free, i.e. they are described by Schr\"odinger equation (\ref{nmain}) with kinetic term only.  
However, the highly nontrivial action of the same operator on holomorphic wavefunctions gives rise to  a finite number of extra bound
states with nonzero angular momentum. These states do not have any commutative analogues and disappear from our theory in the limit
of vanishing $\theta$ as well, while in the limit of strong noncommutativity the spectrum of these states coinsides with the spectrum 
of a harmonic oscillator with frequency $\frac{2}{\theta}$.   

\item
For Dirac particles we can use the correspondence between Fock space operators and ordinary functions
\begin{equation}
|n\rangle\langle m| \sim \frac{\bar z^n}{\sqrt{n!}}\frac{z^m}{\sqrt{m!}}\ e^{-\frac{1}{\theta}r^2}
\end{equation}  
to show that commutative limit of our solutions (\ref{solution1}),(\ref{solution2}) for critical value of $-1<\nu<0$ is
\begin{equation}
\chi_{\nu}(r) \sim \left(
\begin{array}{c}
\sqrt{E+m}J_{-\nu}(kr)\\
-i\sqrt{E-m}J_{-\nu-1}(kr)
\end{array}
\right)
\end{equation}
which after camparison with eq.(\ref{extensions}) tells us that commutative limit of our model corresponds to $\Theta =3\pi/2$ which probably 
explains  the absence of bound states in our model, since in commutative limit bound states exist only if $\pi/2<\Theta<3\pi/2$.

\item
The simple ansatz used to find vector potential is valid only if condition  $1 - \frac{e\Phi}{2\pi} > 0$ is satisfied. It is not clear at 
present if it is possible to extend our approach to $1 - \frac{e\Phi}{2\pi} < 0$ region.
\end{enumerate}

%%%%%%%%%%%%%%%%%%%%%%%%%%%%%%%%%%%%%%%%%%%%%%%%%%%%%%%%%%%%%%%%%%%%%%%%%%%%%%%%%%%%%%
%%%%%%%%%%%%%%%%%%%%%%%%%%%%%%%%%%%%%%%%%%%%%%%%%%%%%%%%%%%%%%%%%%%%%%%%%%%%%%%%%
\appendix
\section{}

Throughout this paper we work in $2+1$ dimensional flat noncommutative space with usual commutation relations:
\begin{equation}
[x^1, x^2] = i\theta. \label{commut}
\end{equation}  
It is convenient to introduce complex variables $z$ and $\bar z$ 
\begin{equation}
z = x^1 + ix^2, \ \ \ \ \ \bar z = x^1 - ix^2
\end{equation}
so that (\ref{commut}) becomes
\begin{equation}
[z, \bar z] = 2\theta
\end{equation}
and $z, \bar z$ can be thought of as a pair of creation-annihilation operators acting in the space of Fock states 
$\{|0\rangle, |1\rangle, \ldots, |l\rangle, \ldots\}$ as
\begin{subequations}
\begin{eqnarray}
\bar z|n\rangle = \sqrt{2\theta}\sqrt{n+1}|n+1\rangle \\
z|n\rangle = \sqrt{2\theta}\sqrt{n}|n-1\rangle \label{Fock}\\
z|0\rangle = 0.
\end{eqnarray}
\end{subequations}
Algebra of functions on the noncommutative plane is then equivalent to the algebra of linear operators in Fock space.
Derivatives on noncommutative plane are the inner derivations
\begin{equation}
\partial_z = -\frac{1}{2\theta}[\bar z, \ldots], \ \ \ \partial_{\bar z} = \frac{1}{2\theta}[z, \ldots] 
\end{equation} 
while integration is the same as trace of operator
\begin{equation}
\int f(x) \to 2\pi\theta \tr f
\end{equation}

The elements of the algebra of functions on noncommutative space can also be identified with ordinary functions on ${\mathbb R}^2$
through the Weyl-Moyal correspondence
\begin{equation}\label{Weyl}
f(x)\mapsto \hat f(\hat x) = \int d^2p\ f(p)e^{i(p_1 \hat x_1 + p_2 \hat x_2 )}  
\end{equation}
where
\begin{equation}
f(p) = \int \frac{d^2x}{(2\pi)^2} f(x)e^{-i(p_1 x_1 + p_2 x_2 )}
\end{equation}
is the usual Fourier transform of $f(x)$. The product of two functions $f$ and $g$ which corresponds to the product of operators $\hat f\hat g$
is given by the Moyal (or star product) formula
\begin{equation}
f*g(x) = \exp\left[ \frac{i}{2}\theta^{ij}\frac{\partial}{\partial x^1_i}\frac{\partial}{\partial x^2_j}\right]
f(x^1)g(x^2)|_{x^1=x^2=x}
\end{equation}
We also need a generalization of the concept of a $\delta$~-~function to noncommutative space. In usual 
field theory $\delta$~-~functions are used to describe localized sources. But because in noncommutative case the space
is smeared at small distances  we cannot construct a truly localized source. Direct application of transform (\ref{Weyl})
to $\delta^2(x)$ gives an operator which is spread out over all of space. Therefore, the most localized source we can
construct in the noncommutative case is a Gaussian wave packet \cite{GrossNekrasov}
\begin{equation}
\delta_\theta(x) = {1\over{\theta\pi}}e^{-{1\over{\theta}}(x_1^2 + x_1^2)}
\end{equation}
whose transform is 
\begin{equation}\label{Nelta}
\hat\delta = |0\rangle\langle0|
\end{equation}
meaning that noncommutative $\delta$-function is in fact a projection operator onto the Fock space groundstate $|0\rangle$. We 
also note that
$$
\int d^2x \delta_\theta(x) = 1
$$
and in $\theta \to 0$ limit we recover the ussual $\delta$-function.

%%%%%%%%%%%%%%%%%%%%%%%%%%%%%%%%%%%%%%%%%%%%%%%%%%%%%%%%%%%%%%%%%%%%%%%%%%%%%%%%%%%%%%%%%%%%%%%%%%%%%%%%%%%%%%%%%%%%%%%%%%

\end{document}